\documentstyle[prl,twocolumn,aps,graphicx]{revtex}
\begin{document}
\title{Electrical and Mechanical Properties of Twisted Carbon Nanotubes}
\author{Alain Rochefort$^\dagger$ and Phaedon Avouris$^\ddagger$.}
\address{$^\dagger$ Centre de recherche en calcul appliqu\'e (CERCA),
5160 boul. D\'ecarie, bureau 400, Montr\'eal, Qc, Canada H3X 2H9.}
\address{$^\ddagger$ IBM Research Division, T.J. Watson Research
Center, P.O. Box 218, Yorktown Heights, NY 10598, USA.}
\maketitle

\begin{abstract} 
We have evaluated the energies required to twist carbon nanotubes (NTs),  and
investigated the effects of these distortions on their electronic structure and
electrical properties. The computed distortion energies are high, indicating
that it is unlikely that extensive twisting is the result of thermal
excitation. Twisting strongly affects the electronic structure of NTs.
Normally metallic armchair $(n,n)$ NTs develop a band-gap which initially
scales linearly with twisting angle and then reaches a constant value. This
saturation is associated with a structural transition to a flattened helical
structure. The values of the twisting energy and of the band-gap are strongly
affected by allowing structural relaxation in the twisted structures. Finally,
we have used the Landauer-B{\"u}ttiker formalism to calculate the electrical
transport of the metal-NT-metal system as a function of the NT distortion.\\
\end{abstract}

Research on the electronic properties of carbon nanotubes (NTs) has been an
especially  active field in recent years. The reasons for this are twofold: on
one hand they are viewed as potential building blocks of nano-sized electronic
devices \cite{tans,bockrath,martel}. On the other hand, they are ideal
materials to study electrical transport phenomena in  low-dimensional systems,
and to test the theoretical models proposed to explain such phenomena. NTs are
usually thought of as a graphene sheet rolled up to form a compact cylinder
with a radius that can be as small as a few Angstr{\"o}ms \cite{dresselhaus}.
It is obvious, however, that this picture of a NT as a straight, geometrically
and atomically perfect tube is somewhat oversimplified. Images of NTs quite
often reveal structural deformations such as bent \cite{tobias}, twisted
\cite{clauss}, or collapsed \cite{chopra} tubes. These deformations may develop
during growth, deposition and processing, or following an interaction with
surface features such as electrodes, or other NTs. An investigation of the
electronic structure of weakly distorted nanotubes has been performed using a
low energy field theory description \cite{kane,kane2}. Here we aim to study the
distortion energies, electronic properties, and electrical transport of carbon
nanotubes under a wide range of twisting distortions.

The electronic properties of carbon nanotubes are usually discussed in terms of
the $\pi$ electronic structure of the 2D-graphene sheet. The mapping of the
graphene sheet onto a cylindrical surface is specified by a superlattice
translational vector $\vec T=n \vec T_1 + m \vec T_2$ where $\vec T_1 = a
(1,0)$ and $\vec T_2 = a (1/2,\sqrt{3}/2)$, and $a$ is the length of the
primitive translation vector of the graphite lattice.  The pair of indices
$(n,m)$ describes how the NT is wrapped to form the cylinder, and determines
its electrical properties. An $(n,n)$, so called "armchair" tube, has only two
linearly independent Fermi points  $\pm \vec K$, where $\vec K=(k_F,0)$. It is
a metal with two bands intersecting the Fermi energy
\cite{hamada,mintmire,saito}. Thus, when an $(n,n)$ NT is atomically perfect
and undistorted, it is expected to exhibit at low bias a quantized conductance
equal to $4 e^2/h$.  However, even in the case of a perfect NT, a reduction of
the conductance may be observed due to the non perfect transmission at the 
NT-metal electrode contacts.

The low energy theory \cite{kane,kane2} used previously to describe the
electronic structure of distorted carbon NTs considers only the $\pi$-orbitals,
and can be appropriately described by a tight-binding model in which
deformations are taken into account simply by modifying the hopping amplitudes
at the deformation sites. This analysis shows that a smooth bending of the tube
along its axis of rotation has little influence on its electronic properties.
On the other hand, it predicts that a twisting deformation can lead to
substantial modification of the electronic structure.  The low energy theory is
applicable to small deformations. Large deformations induce the mixing of
orbitals, so that  describing the electronic structure changes simply by
modifying the hopping amplitudes is no longer adequate. In previous work
\cite{prb}, we computed the electronic properties of bent NTs. We found that
there is a maximal bending angle above which the low energy theory breaks down
and the transport properties are substantially changed. Here we extend this
work by studying the effect of twisting distortions over a wide range of
twisting angles using a model that takes into account of both the $s$ and $p$
carbon orbitals, as well as allowing the geometrical relaxation of the
distorted structures.

The nanotube model used in the computations contains 948 carbon atoms arranged
in an armchair $(6,6)$ structure. The energetics of the deformations were
determined with molecular mechanics using the TINKER program\cite{tinker} with
a modified MM3 force-field \cite{cpl}. Calculations were also performed using a
tight-binding density functional theory (TB-DFT) method developed by Porezag
{\it et al.} \cite{tbdft}. In both types of calculations the dangling bonds at
the ends of the tubes were saturated by hydrogen atoms. Distorted structures
were generated by two different schemes. In the first, continuous twisting, the
circular plane sections of carbon atoms along the tube axis were rotated
sequentially by an (additive) specified angle. In the second, alternate
twisting, a (non-additive) rotation was performed on every second section. The
molecular mechanics calculations showed that it is less energetically demanding
to continuously twist the tube than to twist it alternatively. Therefore, in
the rest of the paper, we will focus our attention on continuous deformations.
The electronic structures of twisted NTs were determined using the extended
H\"uckel method (EHM) \cite{yaehmop}. EHM gives results similar to those
obtained on extended NTs with more sophisticated methods \cite{jpc}. The energy
band-gap was obtained as the difference between the HOMO and LUMO energies.

To evaluate the transport properties, the two ends (with dangling bonds) of the
NT were bonded to gold electrodes. The computation of the transmission function
was then carried out by using the Landauer-B\"uttiker formalism as described in
detail in ref.\cite{datta}. In this scheme, the retarded Green's function for
the system is given by:

\begin{equation}
G^R=\frac{1}{ES_{NT}-H_{NT}-\Sigma_1-\Sigma_2}
\end{equation}

\noindent
where $H_{NT}$ and $S_{NT}$ are the Hamiltonian and the overlap matrices for
the carbon nanotube, and $\Sigma_{1,2}$ are self-energies describing the
interaction of the NT with the metal contacts. These are given by:

\begin{equation}
\Sigma_j=\tau_j^\dagger g_j \tau_j
\end{equation}

\noindent
with $\tau_j$ describing the coupling between the NT and contact $j$ and
$g_j$ the Green's function of the metal used for the contacts. The total
transmission function from one contact to the other is then given by:

\begin{equation}
T(E) = \textrm{Tr}[\Gamma_2 G \Gamma_1 G^\dagger]
\end{equation}

\noindent
with $\Gamma_j=i(\Sigma_j-\Sigma_j^\dagger)$.

In Figure 1, we plot the potential energy for a continuous twisting deformation
against the twist angle for a constrained $(6,6)$ structure, and for a relaxed
one where only the first two carbon sections at each end of the tube were held
fixed. In the inset, we compare the energies obtained with molecular mechanics
for small angle twisting of the relaxed structures to the energies determined
with the TB-DFT method. A number of observations can be made. First, the energy
required for even a moderate twisting of the tube is very large. At low
twisting angles (inset) a quadratic law is obeyed. The twisting energies 
calculated using the MM3 potentials are lower than those computed using TB-DFT
(inset). Therefore, in the following we will consider the MM3 energies as lower
limits to the energetics of twisting deformations.  Figure 1 shows clearly that
structural relaxation  has a very strong effect on the potential energy of
deformation, especially for large twisting angles where the relaxation energy
is computed to be of about the same order of magnitude as the final energy.
Furthermore, the energy profile of the relaxed structure deviates from the
quadratic law (dot-dashed line) for $\Theta_T >$ 14$^{\circ}$nm$^{-1}$. This
deviation corresponds to the onset of a structural transformation where the
tube flattens and takes on a helical shape (such as in Figure 2 [c,d]). A
similar transition has been observed in molecular dynamics simulations
\cite{yakobson}. In order to determine the effect of the diameter on the
collapsing of the nanotubular structures, we have also computed MM3 energies of
twisted $(10,10)$ models. Taking into account the differences in the number of
carbon atoms involved ($(10,10)/(6,6)$ = 1.67), and the NT diameters
($\phi_{(10,10)}/ \phi_{(6,6)}$ = 1.67), we find that the energy needed to
twist a $(10,10)$ NT is similar to that needed to twist a $(6,6)$ NT. The
collapse of the $(10,10)$ tubular structure into the helical shape occurs at a
lower twisting angle than the $(6,6)$ structure, i.e. when $\Theta_T
>10^{\circ}$nm$^{-1}$. However, due to its larger diameter, the relative 
displacement of the carbon atoms during the tube to helix transition for a
$(10,10)$ NT is larger than in the case of a $(6,6)$ NT.

\begin{figure}[tbh]
\centerline{\includegraphics[width=6.5cm]{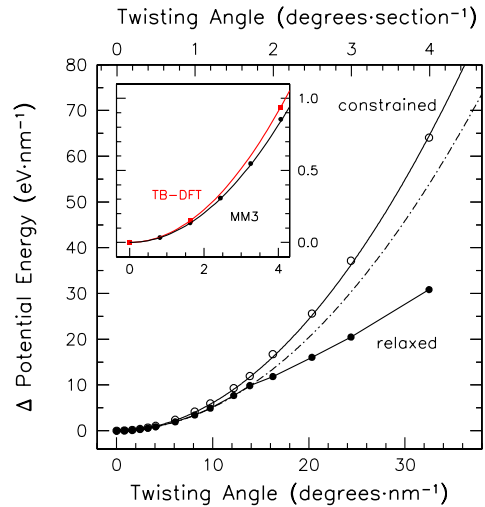}}
\caption{Molecular mechanics calculation of the energy required
to twist a (6,6) armchair NT as a function of twist angle,
for a constrained (open circles) and a relaxed geometry (closed circles). 
The dot-dashed line corresponds to the quadratic law determined
from the low twisting angle region of the relaxed NT. The inset shows 
a comparison of the MM3 and TB-DFT energies for small twisting angle.}
\end{figure}

Our calculations also showed that the twisting energies scale nearly
linearly with the length of the nanotube. Therefore, we can extrapolate and
obtain the twisting energies of longer nanotube segments, and use these
energies to evaluate their Boltzmann distributions as a function of
temperature. The relative population of twisted $(6,6)$ structures can be
described by:

\begin{equation}
F(\Theta_T) = \textrm{exp}\big[ -\alpha \cdot \frac{L{\Theta_T}^2}{T}\big]
\end{equation}

\noindent
where, $\alpha$ is a constant equal to 580 nm$\cdot$degree$^{-2}\cdot$K, $L$ is the
nanotube length in nm, $\Theta_T$ is the twisting angle (degree$\cdot$nm$^{-1}$), and
$T$ is the temperature. From this equation, it is clear that large twists
cannot be generated by thermal excitation. For example, to obtain by thermal
excitation a relatively small population (1\%) of a 1 $\mu$m long $(6,6)$
nanotube twisted by 9.8$^{\circ}$nm$^{-1}$, which is equivalent to the twist
deformation obtained by Clauss {\it et al.} from STM images of nanotube
ropes\cite{clauss}, it would require a temperature larger than $10^6$ K! We
conclude then that large scale nanotube twists must be the result of either
mechanical interactions taking place during the deposition or manipulation of
the nanotube sample, or they are introduced during the high energy  growth
processes and frozen in place by shear forces due to the interaction with the
substrate and other tubes. Local twisting associated with tube collapse is,
however, less energetically demanding and is encountered quite often in AFM
images.

\begin{figure}[tbh]
\centerline{\includegraphics[width=7cm]{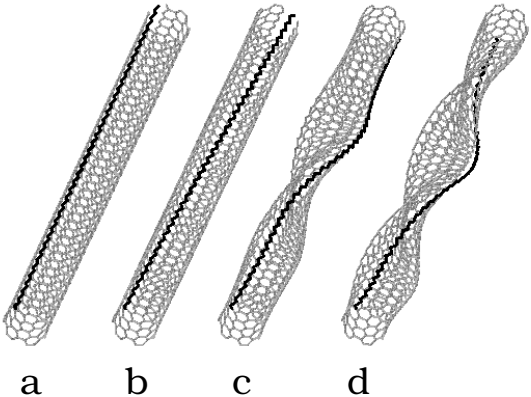}}
\vspace{0.2cm}
\caption{Relaxed NT structures obtained with force-field MM3 energy 
minimization on 0 (a), 1 (b), 2 (c) and 3~$^{\circ}$section$^{-1}$ (d)
twisted models.}
\end{figure}

We now consider the electronic structure changes induced by twisting the NTs.
Figure 3 shows the variation of the band-gap value of a twisted $(6,6)$
nanotube as a function of the twisting angle, for both constrained and relaxed
NT structures.  We find that for the constrained structure, the band-gap
increases linearly with twisting angle up to 16 $^{\circ}$nm$^{-1}$, then
gradually decreases to zero at higher angles. Allowing relaxation in the 
twisted nanotube structure has a strong influence on the value of the band-gap;
the gap increases linearly (but with a different slope) with twisting angle up
to 14 $^{\circ}$nm$^{-1}$, then reaches a stable value ( $\approx$ 0.3 eV) at
higher angles. The first, linear, region corresponds to the opening of a gap
produced by the strong effects of symmetry breaking on the frontier $\pi$ and
$\pi^*$ orbitals. This initial linear increase obtained in constrained
structures is in general agreement with the predictions of low energy theory
\cite{kane} which, however, does not take relaxation into account. For higher
twisting angles, the band-gap decrease observed for the constrained geometry is
mainly due to the presence of highly destabilized $\sigma-\sigma^*$-orbitals
near the Fermi level that emerge from the increasing
$(sp)_{\sigma}-(sp)_{\sigma}$ overlap. The plateau observed in relaxed
structures is the result of the collapse of the tubular structure into a
flattened helix shape which decreases the large repulsive overlap of strained
NTs (Fig.2c,d).

The opening of a band-gap in originally metallic nanotubes leads to a drastic
modification of the electrical properties of nanotubes upon twisting by even
low angles. In order to investigate the conduction of electrons in twisted
(and relaxed) $(6,6)$ nanotubes, we have connected both tube ends (with
dangling bonds on the end carbons) to gold electrodes. Each electrode is
composed of a layer of 22 gold atoms in a (111) crystalline arrangement. The
distance between the NT end and the gold layer is 1.0 {\AA}. Such a bonding
configuration minimizes the contact-resistance and emphasizes the transport
properties of the nanotube itself. The Hamiltonian ($H_{NT}$) and overlap
matrices ($S_{NT}$) of equation (1) are determined using EHM for the system
Gold-NT-Gold. The transmission function $T(E)$, that represents the sum of
the transmission probabilities over the contributing nanotube conduction
channels, is then computed from equation (3). The $T(E)$ spectra are expressed
with respect to the Fermi level ($E_F$ is defined as 0 eV) of individual
twisted nanotubes.

\begin{figure}[tbh]
\centerline{\includegraphics[width=6.5cm]{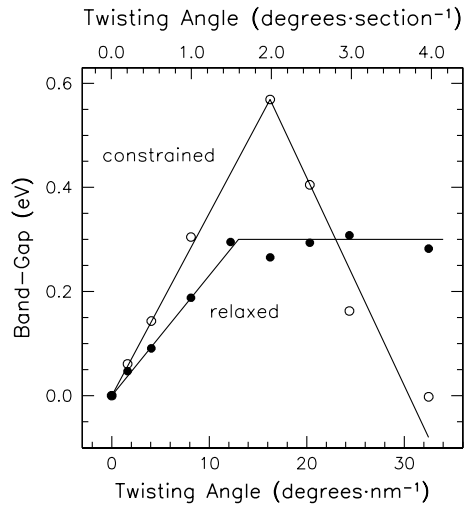}}
\caption{Band-gap opening in a (6,6) nanotube as a function of the twist angle 
for a constrained geometry (open circles), and a relaxed geometry (closed 
circles).}
\end{figure}

Figure 4 shows the transmission $T(E)$ spectra of nanotubes twisted by 0, 1.6,
4.1 and 8.1$^{\circ}$nm$^{-1}$ angles (main panel). All these deformed tubes
lie within the linear regime of the band-gap variation with twist angle. The
variation of $T(E)$ at $E_F$ with twisting angle is shown at the upper-left
panel, while the corresponding change in resistance at zero bias (= 12.9
k$\Omega$/$T(E_F)$) is given at the upper-right panel. The spectrum of the
perfect structure (0$^{\circ}$nm$^{-1}$) shows a $T(E)$ at the Fermi level of
about 1.2, leading to a resistance higher than expected for ballistic transport
(where $T(E)$ = 2.0). This reduction in transmission is due to the presence of
a finite contact resistance. The increasing $T(E)$ on the high binding energy
side is due to the opening of higher conduction channels.  The asymmetry of the
transmission function T(E) above and below $E_F$ depends on the NT-metal
electrode coupling (Au-NT distance). A larger Au-NT distance leads to a higher
T(E) above $E_F$ , and vise versa. Since the NT-metal electrode distance is
kept fixed in all calculations, this effect of contact  resistance does not
influence the effects induced by tube twisting.  

\begin{figure}[tbh]
\centerline{\includegraphics[width=7.5cm]{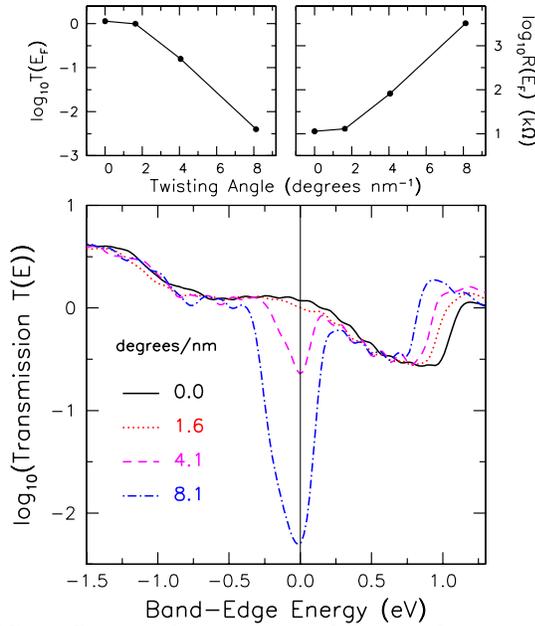}}
\caption{Computed transmission function of a twisted (6,6) NT. The upper-left
inset shows the logarithmic decay of $T(E_F)$ with twisting angle, and
the upper-right inset gives the variation of the resistance at zero bias
for the same twisted NT.}
\end{figure}

Twisting induces important changes in the $T(E_F)$, a property that determines
the linear conductance of the system.  Already a twist of 4$^{\circ}$ nm$^{-1}$
(or $0.5^{\circ} $section$^{-1}$), reduces the transmission  by a factor of 14.
It is interesting, however, that essentially no change is seen in $T(E_F)$ at
low twisting angles ($< 2^{\circ}$nm$^{-1}$) despite the fact that  a finite
band-gap ($\approx 0.04$ eV) is calculated for the free NT. We attribute this
behavior to the metal-induced gap states (MIGS)  produced by the interaction of
the short (96 {\AA}) NT segment with the gold metal electrodes \cite{heine} and
the presence of charge-transfer doping. The MIGS tend to bridge the
distortion-induced gap when the latter is small \cite{leonard}. For twisting
angles larger than $2^{\circ}$nm$^{-1}$, the transmission decreases
exponentially (see upper-left in Fig. 4). At 8$^{\circ}$nm$^{-1}$, $T(E_F)$ has
been decreased by  more than 2 orders of magnitude, and if we consider
transport into a strongly  twisted, helical shape nanotube (see Fig.2 [c,d]),
the decrease in $T(E)$ is about 5 orders of magnitude. A twisting-induced
band-gap is, most likely, responsible for the recently documented field-effect
transistor action of a twisted, normally metallic multi-wall nanotube
\cite{martel}.

In contrast to our findings on axially bent nanotubes\cite{prb} where for a
$(6,6)$ tube the decrease in transmission occured at energies below $E_F$,
and was traced to curvature-induced localized $\sigma$-$\pi$ orbital
mixing, the sharp decrease in $T(E)$ upon twisting straddles $E_F$ and is
induced by symmetry breaking all along the NT.\\

We thank Dr. Thomas Heine for providing us the tight-binding
DFT code, and Dr. Fr\'ed\'eric Lesage for very helpful discussions.

\end{document}